\documentclass[aps,prl,preprint,showpacs,letterpaper,superscriptaddress]{revtex4}
\usepackage{graphicx}
\usepackage{amsmath} 
\usepackage{grffile}
\usepackage{color}
\def\beqn{\begin{eqnarray}}
\def\eeqn{\end{eqnarray}}
\def\beqns{\begin{eqnarray*}}
\def\eeqns{\end{eqnarray*}}
\def\beq{\begin{equation}}
\def\eeq{\end{equation}}
\def\bea{\begin{array}}
\def\ea{\end{array}}

\def\<{\langle}
\def\>{\rangle}

\newlength{\textlarg}

\newcommand{\be}{\begin{equation}}
\newcommand{\ee}{\end{equation}}
\newcommand{\ben}{\begin{eqnarray}}
\newcommand{\een}{\end{eqnarray}}
\newcommand{\B}{\mathrm{B}}
\newcommand{\NB}{\mathrm{nB}}

\DeclareGraphicsExtensions{.ps,.eps,.EPSF} 

\begin{document}
\title{Electronic structure and transport in approximants of the Penrose tiling}
\author{Guy \surname{Trambly de Laissardi\`ere}}
\affiliation{Laboratoire de Physique th\'eorique et Mod\'elisation, CNRS and 
Universit\'e de Cergy-Pontoise, 
F-95302 Cergy-Pontoise, France}
\author{Attila \surname{Sz\'all\'as}}
\affiliation{
Wigner Research Centre for Physics, P.O. Box 49, H-1525 Budapest, Hungary
}
\author{Didier \surname{Mayou}}
\affiliation{
Univ. Grenoble Alpes, Institut NEEL, F-38042 Grenoble, France \\
CNRS, Institut NEEL, F-38042 Grenoble, France
}
\date{\today}

\begin{abstract}
We present numerical calculations of electronic structure and transport  in Penrose approximants.
The electronic structure of perfect approximants  shows a spiky density of states and a tendency to localization that is more pronounced in the middle of the band. Near the band edges the behavior is more similar to that of free electrons. These calculations of  band structure and in particular the band scaling suggest an anomalous quantum diffusion when compared to normal ballistic crystals. This is confirmed by a numerical calculation of quantum diffusion which shows a crossover from normal ballistic propagation at long times to anomalous, possibly insulator-like, behavior at short times. The time scale $t^*(E)$ for this crossover is computed for several approximants and  is detailed. The consequences for electronic conductivity are discussed in the context of the relaxation time approximation. The metallic like or non metallic like behavior of the conductivity is dictated by the comparison between the scattering time due to defects and the time scale $t^*(E)$. 
\end{abstract}

\maketitle

{\bf 1. Introduction}

\vskip .5cm
Since the discovery of  Shechtman et al. \cite{shechtman84}
numerous experimental studies indicated
that the conduction
properties of several stable quasicrystals (AlCuFe, AlPdMn, AlPdRe...) are quite opposite to those of
good metals
\cite{Poon92,Berger93c,Belin93,Berger94,Grenet00_Aussois}.
It appears also that the medium range order, over one or a few nanometers, is the relevant length scale
that determines conductivity. 
In particular, the role of transition elements enhancing localization has been often studied 
\cite{Fujiwara89,Fujiwara93,Krajvci95,GuyPRB97,GuyICQ6,PMS05}.
There is now strong evidence that these nonstandard properties
result from a new type of breakdown of the semiclassical Bloch-Boltzmann theory of conduction
\cite{PRL06,Trambly08,Trambly14CRAS,Mayou08RevueTransp}.
On the other hand, the specific role of long range quasiperiodic order on transport properties is
still an open question in spite of a large number of studies (see Refs.  
\cite{Kohmoto86,Fujiwara89Fibo,Sire90,Passaro92,Sire94,Jagannathan94,Yamamoto95,Fujiwara96,Zhong94,Roche97,Mayou00,Zijlstra00,Zijlstra00b,Triozon02,Zijlstra02,Zijlstra04,Jagannathan07,Trambly11} 
and Refs. therein).
In this paper, we study ``how electrons propagate''
in approximants of 
the rhombic Penrose tiling P3 (PT in what follows).
This tiling is one of the well-known  quasiperiodic tilings
that have been used to understand the influence of quasiperiodicity on electronic transport
\cite{Sire90,Passaro92,Jagannathan94,Zhong94,Zijlstra02,Zijlstra04,Jagannathan07}.
The main objective is to show that non standard conduction properties result from
purely quantum effects due to quasiperiodicity that cannot be interpreted
through the semiclassical theory of transport.

\vskip .5cm
{\bf 2. Approximants of Penrose tiling}

\vskip .5cm
To study electronic properties of Penrose tiling, we consider a series of periodic approximants, called Taylor approximants, proposed by M. Duneau and M. Audier \cite{Duneau94}. 
These approximants have defects as compared to the infinite perfect tiling, but the relative number of defects becomes negligible as their size increases. 
They has been used to study the magnetic properties of PT \cite{Szallas07,Szallas08}.
Here we study electronic structure and quantum diffusion in three Taylor approximants, 
${\rm T} = 3$, 4 and 5. Their rectangular cells 
$ L_x\times L_y $ are 
$24.80a \times 21.09a $,  
$40.12a \times 34.13a $, and
$64.92a \times 55.23a $, 
respectively. $a$ is the tile edge length.
They contain 644, 1686 and 4414 sites, respectively.

\vskip .5cm
{\bf 3. Electronic structure}
\vskip .5cm

We study a pure hopping Hamiltonian
\ben
  \hat{H} ~=~  \gamma ~ \sum_{\langle i,j \rangle} | i  \rangle \langle j |
\label{Eq_pureHopHamilt}
\een
where $i$ indexes $s$ orbitals $| i  \rangle$ located on all vertexes. 
For realistic order of magnitude of the model one can choose the strength of the hopping between orbitals $\gamma = 1\,$eV.
Indices $i$, $j$ label the nearest neighbors 
at tile edge distance $a$. 
The properties of this model depend only on the topology of the
tiling.
The electronic eigenstates $| n \vec k  \rangle$, with wave vector $\vec k$ and energy $E_n(\vec k)$, are computed by diagonalization in the reciprocal space for a number $N_k$ of vectors $\vec k$ in the first Brillouin zone. 
The density of states (DOS), $n(E)$, is calculated by,
\begin{equation}
n(E) = \left \langle \delta (E- \hat{H}) \right \rangle_{E_n=E} 
\end{equation}
where  $\left \langle ... \right \rangle_{E_n=E}$ is the average on states with energy $E$.
It is obtained by taking the eigenstates for each
$\vec{k}$ vector with energy $E_n(\vec{k})$ such that
$E-\delta E /2<E_n(\vec{k})<E+\delta E/2$.
$\delta E$ is the energy resolution of the calculation. 
When $N_k$ is too small, the calculated
quantities are sensitive to  $N_k$.
Therefore $N_k$ is increased
until the results do not depend significantly on $N_k$.
We use $\delta E = 0.01$\,eV;
$N_k = 144^2$, $96^2$ and $48^2$ 
for  Taylor approximants ${\rm T} = 3$, 4 and 5, respectively.

\begin{figure}[]
\begin{center}
\includegraphics[width=8cm]{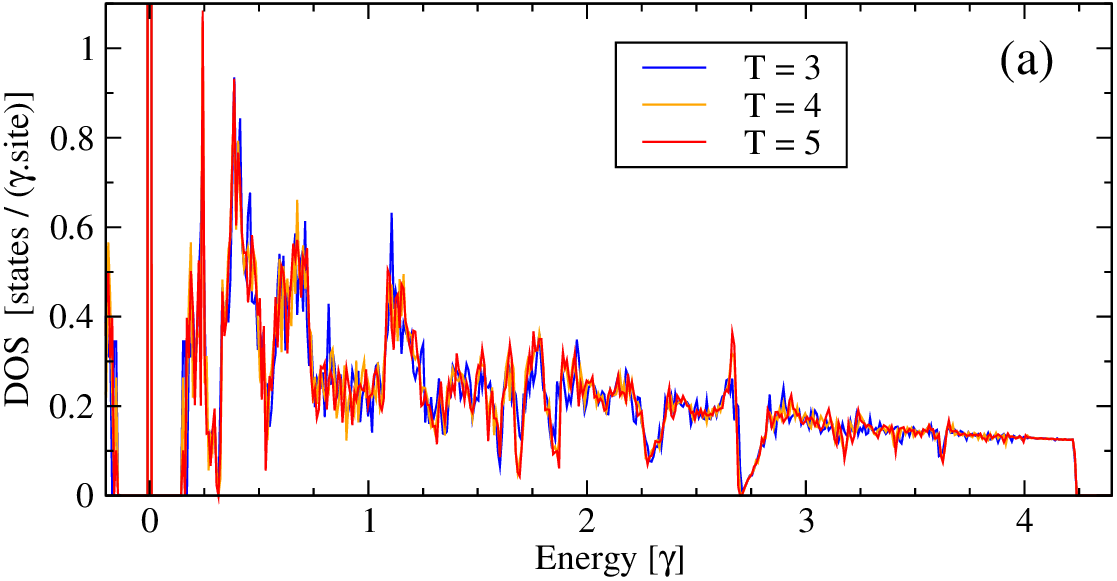}

\vskip .1cm
\includegraphics[width=8cm]{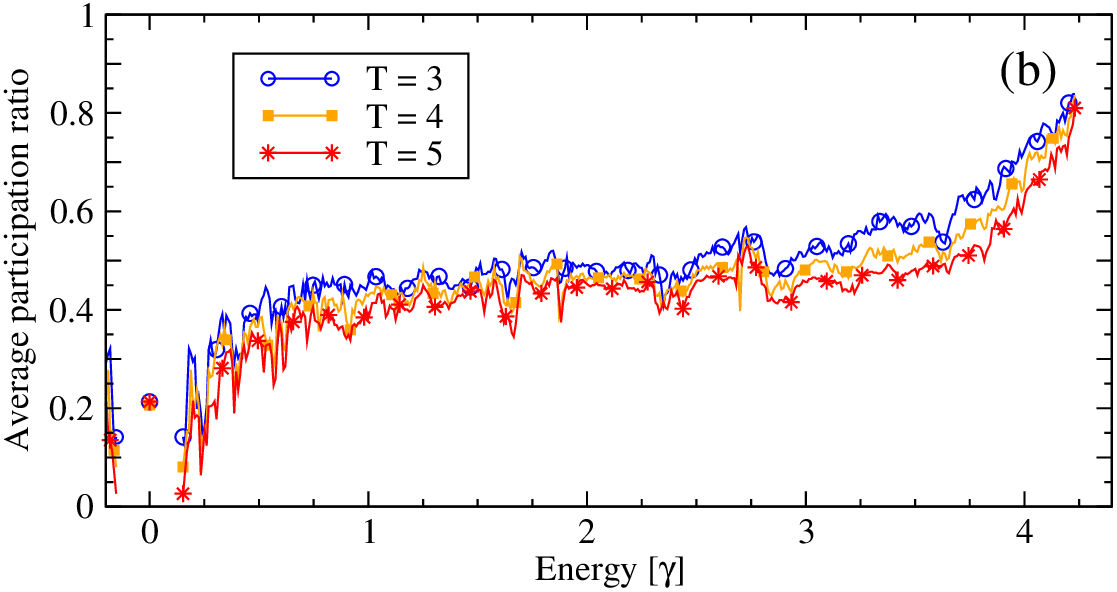}

\vskip .1cm
\includegraphics[width=8cm]{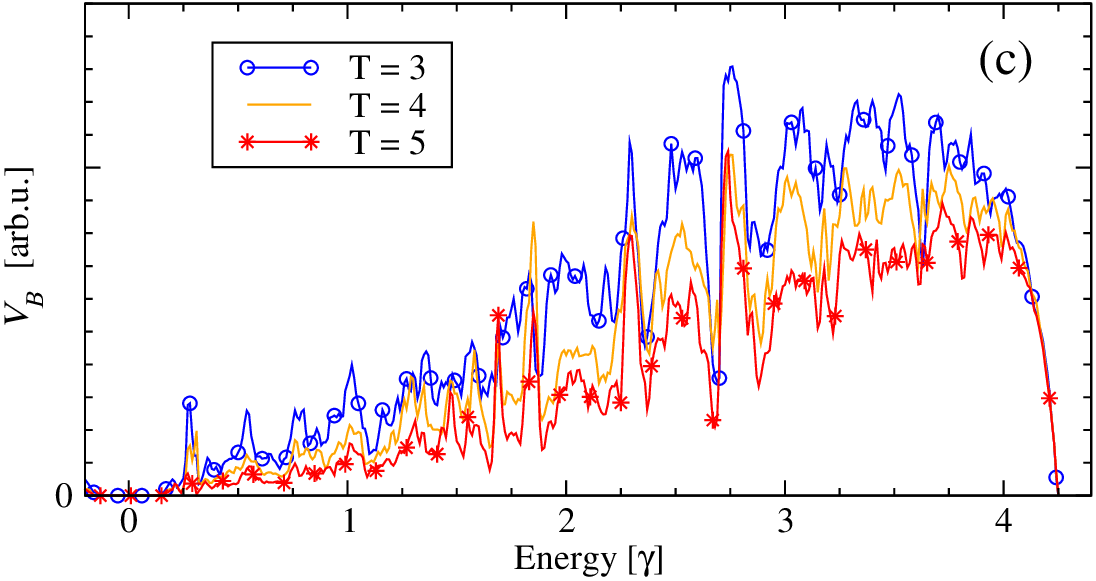}

\vskip .1cm
\includegraphics[width=8cm]{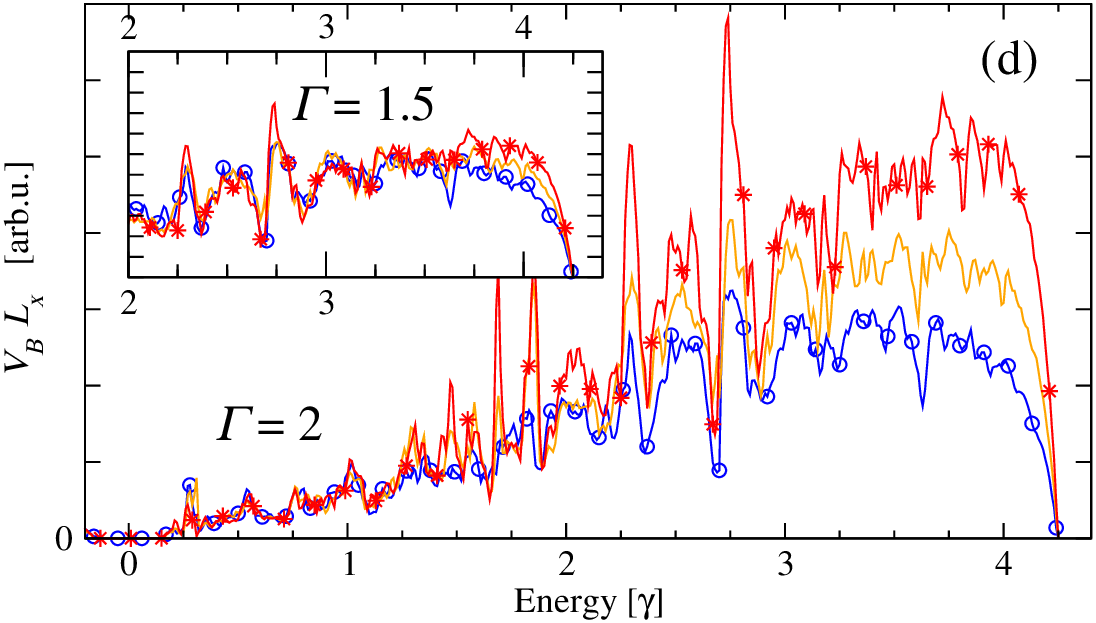}

{Fig. 1: 
(colour online)
Electronic structure in Penrose approximants.
(a) 
Total density of states (DOS) $n(E)$.
DOS is symmetric w.r.t. $E=0$.
(b) Average participation ratio  ${p}(E)$.
(c) Average Boltzmann velocity $V_{\rm B}(E)$ along the $x$ direction,
(d) $ V_{\rm B}(E) \times L_x^{{\it \Gamma}-1} $ versus energy $E$ for ${\it \Gamma} = 2$
[Insert: ${\it \Gamma} = 1.5$.]
}
\end{center}
\end{figure}

\vskip .5cm
{\it Density of states}
\vskip .2cm

The density of states is shown in figure 
1a. 
As expected \cite{Kohmoto86,Zijlstra00,Zijlstra00b}, it is symmetric with respect to $E=0$. 
The main characteristic of these DOS are 
similar to that obtained by Zijlstra \cite{Zijlstra00,Zijlstra00b}, 
for an other family of Penrose approximants. 
At $E=0$ a strictly localized state is obtained \cite{Kohmoto86,Arai88}. 
A gap is found for energy $|E| < \sim 0.13$\,eV and a small gap with a 
width less than 0.01\,eV seem to be at $|E| \simeq 2.7$\,eV \cite{Zijlstra00,Zijlstra00b}. 
Other fine gaps could be present at $ |E| \simeq0.3$, 0.5, 1.7\,eV (...) but 
our energy resolution can not obtain them.
The DOS is more spiky
at the center of the band ($|E|< 2$) and smooth
near the band edges ($|E| > 2$).

\vskip .3cm
{\it Participation ratio}
\vskip .2cm

In order to quantify this localization phenomenon, we
compute  the average participation ratio
defined by,
\ben
{p}(E) = \left \langle  \Big( N \sum_{i=1}^N |\langle i | n\vec k \rangle|^{4} \Big)^{-1} \right \rangle_{E_n=E},
\een
where 
$i$ indexes orbitals in a unit cell
and $N$ is the number of atoms in this unit cell. 
For completely delocalized eigenstates $p$ is equal to $1$.
On the other hand, states localized on one site have a small
$p$ value: $p=1/N$.
Figure 1b 
shows clearly a stronger localization of electronic states for larger
approximants.

\vskip .3cm
{\it Band scaling}
\vskip .2cm

The average Boltzmann velocity along the $x$ direction
is computed by,
\be
V_{{\rm B}}(E) =
\sqrt{
\left\langle 
|\langle n\vec k | \hat{V}_x | n\vec k \rangle |^2
\right\rangle_{E_n=E}
}\, , 
\label{equationVB}
\ee
where the velocity operator along the $x$ direction is $\hat{V}_x = [\hat{X},\hat{H}] /(i \hbar)$, with $\hat{X}$ the position operator.
$V_{\rm B}$ is the average intra-band velocity, 
\be
V_{\rm B}(E) = \frac{1}{\hbar} \left \langle \frac{\partial E_n(\vec k)}{\partial k_x} \right \rangle_{E_n=E} .
\label{equationVB_2}
\ee
Figure 1c 
shows a smaller velocity at the center of the band ($|E|<2$).
When the size of the approximant increases, $V_\B$ decreases as expected from band scaling
analysis \cite{Sire90,Passaro92,Sire94,Mayou08RevueTransp}.
Typically the width $\Delta E$ of a band $E_n(\vec k)$ varies in the $k_x$ direction like, 
$\Delta E \propto L_x^{-{\it \Gamma}}$, where $L_x$ is the length of the unit cell in the $x$ direction. 
The exponent $\it \Gamma$ depend on $E$ and the diffusion properties of the structure. 
For normal metallic crystals ${\it \Gamma} = 1$,
for disordered metallic alloys the electronic states are diffusive and ${\it \Gamma} = 2$. 
From equation (\ref{equationVB_2}), the Boltzmann velocity should satisfy that
$V_{\rm B} \propto L_x^{1-{\it \Gamma}}$.
Figure 1d 
shows $V_{\rm B} L_x^{{\it \Gamma}-1}$ versus energy $E$.
For ${\it \Gamma} \simeq 2$ the value of $V_{\rm B}(E) L_x^{{\it \Gamma}-1}$ are rather similar for the
three approximants at the center of the band ($|E|<2$).
For $2<|E|< 3.5$, it seems that ${\it \Gamma} \simeq 1.5$, and near the band edges, $|E|>3.5$, states 
are almost ballistic ${\it \Gamma} \simeq 1$.

\vskip .5cm
{\bf 4. Electronic transport}

\vskip .5cm
{\it Quantum diffusion}
\vskip .2cm

%

\begin{figure}[]
\begin{center}

~\includegraphics[width=8cm]{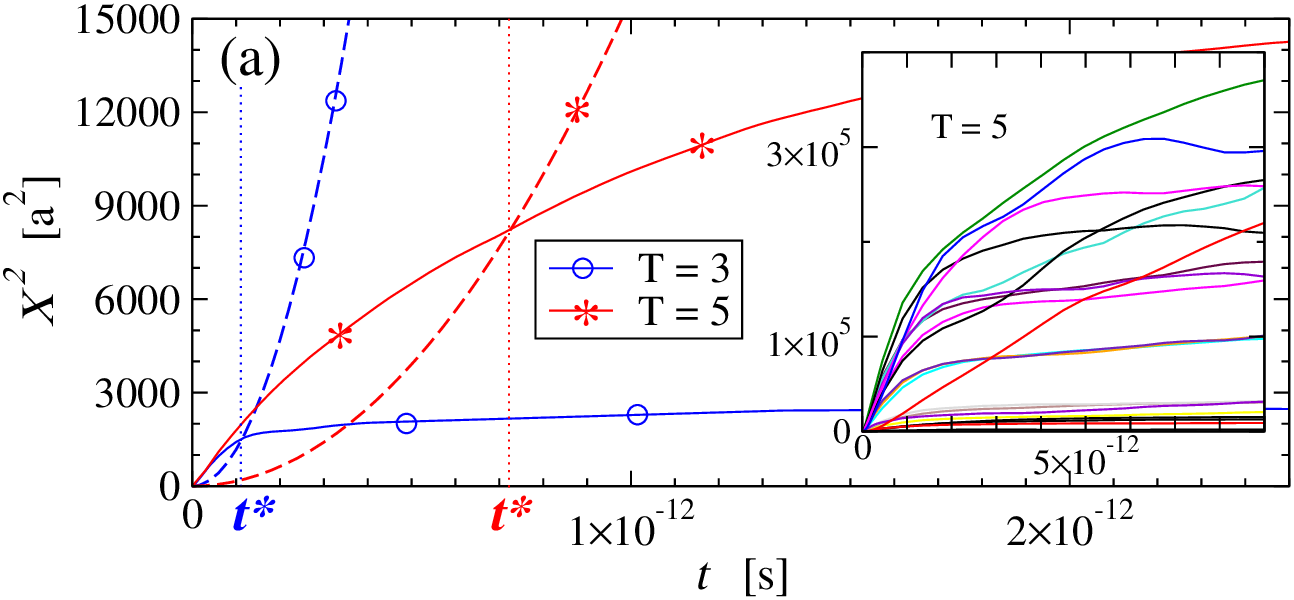}~~

\vskip .2cm
\includegraphics[width=8cm]{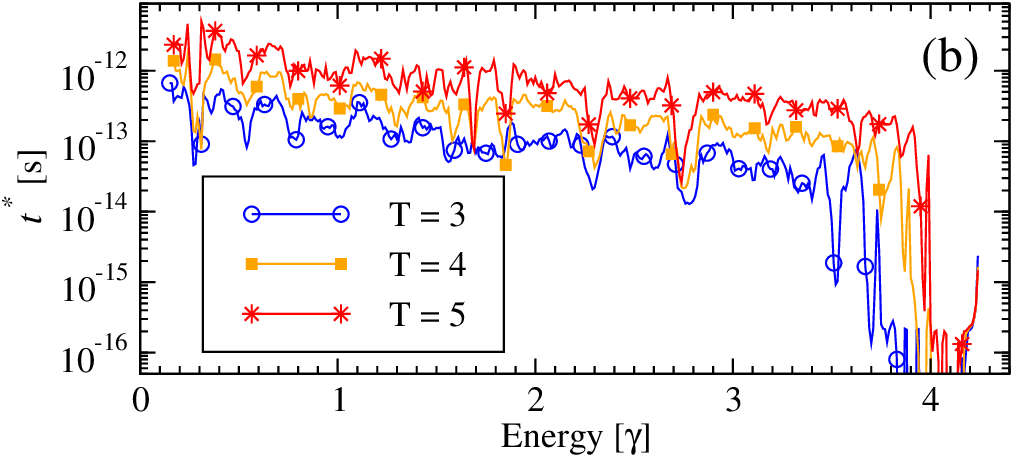}
\end{center}

{Fig. 2: 
\label{Fig_X2_tetoile}
(colour online)
(a) Average square spreading versus time $t$
at $E=1.0$\,eV
in perfect Penrose approximants $\rm T=3$ and 5: (dashed line) Boltzmann $X^2_{\B}$ and (line) Non-Boltzmann 
$ X^2_{\NB}$.
[Insert: $X^2_{\NB}$ versus time $t$ at different energies in approximant $\rm T = 5$.]
(b) Time $t^*$
in perfect Penrose approximants (see text).
}
\end{figure}

The band scaling has a direct consequence for the 
wave propagation in the medium.
The mean spreading,
$L_{\rm wp}(t)$ of a wavepacket is neither ballistic (i.e. proportional to time $t$) as in perfect crystals nor diffusive (i.e. $L_{\rm wp}(t) \propto \sqrt{t}$) as in disordered metals. In general  at large $t$,
\be
L_{\rm wp}(E,t) \propto t^{\beta(E)}
\ee
The value of the exponent $\beta$ in quasicrystals (or in approximants with size cell $L_x$ going to infinity) can be related to ${\it \Gamma}$ in finite approximants by  $\beta  = 1/{\it \Gamma}$ \cite{Mayou08RevueTransp}.  
Thus our results on approximants show that  states in Penrose tiling 
are diffusive ($\beta \simeq 0.5$) at the center of the band ($|E|<2$),
super-diffusive ($0.5<\beta< 1$) for $2<|E|<3.5$,
and almost ballistic ($\beta \sim 1$) near the band edges ($|E|>3.5$). 

It is possible to go beyond these qualitative arguments by defining in an exact manner the quantum diffusion as we show now. The average
square spreading of states of energy $E$ at time $t$
along the $x$ direction, is defined as: 
\begin{eqnarray}
 X^{2}(E,t) =\Big< \Big(\hat{X}(t)-\hat{X}(0) \Big)^{2}\Big>_{E} ,
 \label{Def_2}
 \end{eqnarray}
with $\hat{X}(t)$ the Heisenberg representation of the operator. It can be shown that $\hat{X}$,
is the sum of two term \cite{PRL06,Trambly14CRAS},
\begin{equation}
 X^2(E,t) = V_{\B}^2(E) t^2 + X_{\NB}^2(E,t).
\label{eqL2BetNB}
\end{equation}
The first term, $ X_{\B} = V_{\B}^2(E) t^2$, is the ballistic (intra-band)  contribution
at energy $E$. 
The semiclassical model of 
the Bloch-Boltzmann transport theory amounts to
taking into account only this first term.
The second term (inter-band contributions)
$X^2_{\NB}(E,t)$  is
a non-Boltzmann contribution.
It is due to the non-diagonal
elements of the velocity operator and describes
a spreading of the wave-function.
 
One defines the time $t^*(E)$ for which 
$X_{\B}^2 = X_{\NB}^2$ at energy $E$ 
(figure 2a). 
For long time, $t > t^*$, the ballistic semiclassical contribution
dominates the quantum diffusion but for short time,  $t < t^*$, the non-ballistic
contribution dominates (``low velocity regime'' \cite{PRL06}). Therefore $t^*(E)$ is an important time scales for any approximant. On time scale larger than $t^*(E)$ the approximant behaves like a  normal metal whereas on smaller time scales the 
approximant may behave quite differently and, in particular,  may show insulator-like behavior
(see below). 
When the size of the approximants increases, the characteristic time limit
$t^*$ of the crossover
between ballistic and non-ballistic behavior increases 
(figure 2b.) 
Similar results has been found in approximants of octagonal tiling \cite{Trambly11}. From
ab-initio electronic structure calculation and in realistic approximants  $\alpha$-AlMnSi \cite{PRL06}, 1/1 AlCuFe  \cite{Trambly08} and in the complex metallic hexagonal phase $\lambda$-AlMn \cite{Trambly14CRAS}  it has been shown that the order of magnitude of $t^*(E)$ is about $10^{-14}$ or $10^{-13}$\,s.

$X^2_{\NB}(E,t)$ oscillates and is bounded by 
$L_{\NB}(E)^2$, which depends on the energy $E$ \cite{Trambly14CRAS}.
From numerical calculations 
(figure 2a), 
it is found that for many energies $E$,
$X^2_{\mathrm{\NB}}(E,t)$ reaches rapidly its maximum limit $L_{\NB}(E)^2$ 
and one can assume, $X_{\NB}(E,t)^2 \simeq L_{\NB}(E)^2$
for large $t$, and $t^* \simeq L_{\NB}/V_{\B}$.

\vskip .3cm
{\it Conductivity in the Relaxation Time Approximation}
\vskip .2cm

\begin{figure}[]
\begin{center}
\includegraphics[width=7cm]{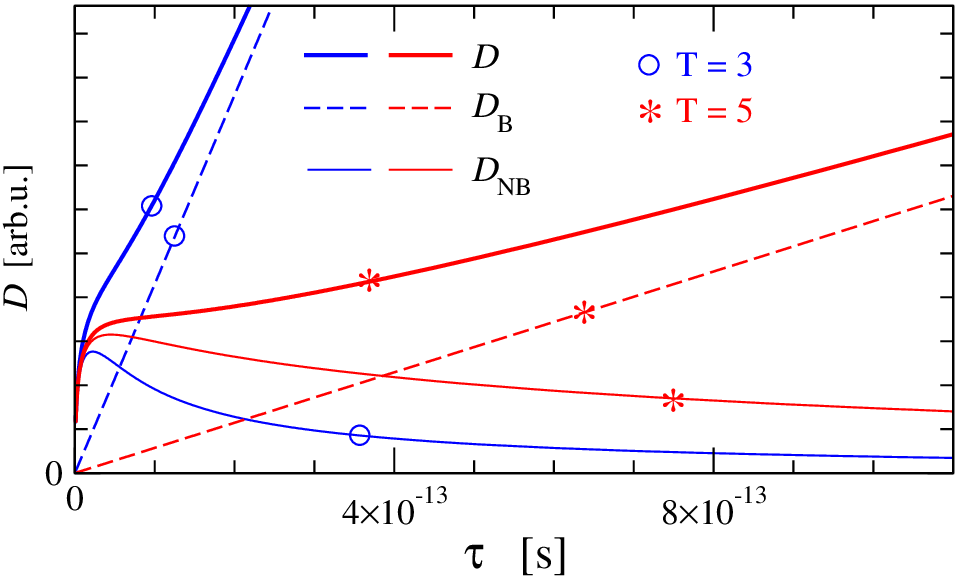}

{Fig. 3: 
(colour online)
Diffusivity $D$, $D = D_{\B} + D_{\NB}$, in relaxation time approximation, versus scattering time $\tau$,
at $E_{\rm F}=1.0$\,eV
in Penrose approximants $\rm T=3$ and 5.
}
\end{center}
\end{figure}

In the Relaxation Time Approximation (RTA) \cite{Mayou00,PRL06,Mayou08RevueTransp}
the role of phonons and static defects are taken into account through a scattering time $\tau$.
$\tau$ decreases when temperature $T$ increases and when the number of static defects increases.
The scattering time estimates in quasicrystals and approximants from transport measurements at low temperature (4\,K) \cite{Poon92,Berger93c} is about a few  $10^{-14}$\,s or even more.
That is close to the time limit $t^*$ between Boltzmann and non-Boltzmann behavior (see previous section).
Therefore  the non-Boltzmann behavior could play a crucial role in the conductivity.

At zero temperature, the static conductivity is given by the Einstein formula,
\ben
\sigma = e^2 n(E_{\rm F}) D(E_{\rm F})\, ,
\een
where $E_{\rm F}$ is the Fermi energy 
and $D$, the diffusivity which is the sum of a Boltzmann and non-Boltzmann terms,\cite{PRL06}
\begin{equation}
D(E_{\rm F}) = D_{\B} + D_{\NB}(E_{\rm F}), 
{\rm ~~~with~}D_{\B} = V_{\rm B}^2(E_{\rm F}) \tau .
\label{Eq_SVR_D}
\end{equation}
$D_{\NB}(E_{\rm F})$ is calculated numerically from eigenstates (see \cite{Mayou08RevueTransp,Trambly14CRAS}). 
Figure 3 
shows the diffusivity in approximants $\rm T=3$ and 5 for $E_{\rm F} = 1$\,eV.
At very low $\tau$, $\tau< \sim 10^{-15}$\,s,  diffusivity is always ballistic, 
for larger $\tau$ values up to $\tau \simeq t^*$ the non-Boltzmann terms dominate, and for  $\tau \gg t^*$
periodicity of approximants induces ballistic diffusivity.  The intermediate zone, with a non metallic (non ballistic) behavior due to structure, is more important in the largest approximant, and it corresponds to realistic values of scattering time.

\vskip .5cm
{\bf 5. Conclusion}

\vskip .5cm
To summarize, we have presented a numerical study of electronic structure and quantum diffusion for a pure hopping Hamiltonian in approximants 
of Penrose tiling containing 644, 1686 and 4414 sites in a unit cell.
When the size of the unit cell of the approximant increases the usual Boltzmann term for quantum diffusion (ballistic term) decreases rapidly and 
non-Boltzmann terms become essential to understand transport properties. 
These non-Boltzmann terms can have  ``insulator-like'' behavior, suggesting that in larger approximants,  ``insulator-like'' states,
due to long range quasiperiodic order, could exist. Calculations in larger approximants are in progress.

\vskip .5cm
{\bf Acknowledgements}

\vskip .5cm
We thank  A. Jagannathan  for fruitfull discussions.
The computations were performed at the
Centre de Calcul of the 
Universit\'e de Cergy-Pontoise.
We thank Y. Costes and D. Domergue for computing assistance.


\end{document}